\documentclass[aps,prper,superscriptaddress,longbibliography,floatfix]{revtex4-2}

\usepackage{graphicx}
\usepackage{float}
\usepackage{flafter}
\usepackage{booktabs}
\usepackage{amsmath}
\usepackage{array}
\usepackage{xcolor}
\usepackage{pgfplots}
\usepackage{pgfplotstable}
\pgfplotsset{compat=1.18}
\usepgfplotslibrary{groupplots,colormaps}
\usetikzlibrary{calc}
\definecolor{mainblue}{HTML}{1f77b4}
\definecolor{mainorange}{HTML}{e08214}
\definecolor{maingreen}{HTML}{2ca02c}
\pgfplotsset{
  totalscatteraxis/.style={
    xmin=0,xmax=100,ymin=0,ymax=100,
    width=0.217\linewidth,height=0.217\linewidth,
    scale only axis,
    axis lines=left,
    axis line style={black},
    tick style={black},
    tick label style={font=\tiny,black},
    label style={font=\tiny,black},
    title style={font=\scriptsize,black},
    grid=major,
    grid style={gray!20},
    xtick={0,50,100},
    ytick={0,50,100},
    clip=false
  },
  paperbaraxis/.style={
    ybar=0pt,
    /pgf/bar width=10pt,
    axis lines=left,
    axis line style={black},
    tick style={black},
    tick label style={font=\scriptsize,black},
    label style={font=\scriptsize,black},
    title style={font=\small,black},
    grid=major,
    grid style={gray!20},
    enlarge x limits=0.08
  }
}
\newcommand{\scatterstats}[4]{%
  \node[anchor=north west,align=left,font=\tiny,fill=white,fill opacity=0.82,text opacity=1,inner sep=1pt]
    at (rel axis cs:0.03,0.97) {$r=#1$\\$\rho=#2$\\$D=#3\%$\\MAD$=#4\%$};%
}
\usepackage{multirow}
\usepackage{makecell}
\usepackage{placeins}
\usepackage{hyperref}
\hypersetup{colorlinks=true,linkcolor=blue,citecolor=blue,urlcolor=blue}

\newcommand{\barlegenditem}[2]{{\color{#1}\rule{1em}{0.7em}}\hspace{0.3em}\textbf{#2}}
\newcommand{\tablecell}[2][0.64\textwidth]{\parbox[t]{#1}{\raggedright #2\strut}}
\begin{document}

\title{Large-scale AI grading of handwritten physics assessments: Score agreement and Olympiad team selection outcomes}
\author{Praveen Pathak}
\email{praveen@hbcse.tifr.res.in}
\affiliation{Homi Bhabha Centre for Science Education--TIFR, Mumbai, India}
\affiliation{Lawrence Livermore National Laboratory, Livermore, CA, USA}
\author{Siddharth Tiwary}
\email{siddharthtiwary@berkeley.edu}
\affiliation{University of California, Berkeley, CA, USA}
\author{Charudatt Kadolkar}
\affiliation{Indian Institute of Technology, Guwahati, India}
\author{Vijay Singh}
\affiliation{Centre for Excellence in Basic Sciences, Mumbai, India}
\author{David Rakestraw}
\affiliation{Lawrence Livermore National Laboratory, Livermore, CA, USA}
\author{Shirish Pathare}
\affiliation{Homi Bhabha Centre for Science Education--TIFR, Mumbai, India}
\author{Anwesh Mazumdar}
\affiliation{Homi Bhabha Centre for Science Education--TIFR, Mumbai, India}
\date{\today}

\begin{abstract}
Multimodal AI can read handwritten physics solutions, but high-stakes grading requires agreement with official scores and outcomes. This study evaluated GPT-5.5-based grading on $10\,364$ scanned pages from 520 handwritten submissions by 416 unique candidates or students across three assessments: a national Physics Olympiad theory examination, the final Olympiad selection camp with theory and experiment components, and a university quantum-mechanics examination. Each submission was graded twice by AI using the official rubrics. The second round used revised page-by-page and evidence-location instructions developed after first-round disagreement analysis. During grading, AI did not see official human marks or AI--human comparisons. Total-score correlations with official marks were high (0.91--0.97). For the final Olympiad selection, AI recovered the same five-student team as official grading. The second round improved aggregate question-part agreement, especially where first-round disagreements were larger. The main difficulty remained exact partial-credit grading, especially in experimental work. Reliable AI grading therefore depends on detailed rubrics and should be used as a second reader or audit tool under examiner control.
\end{abstract}

\maketitle
\section{Introduction}
Grading handwritten student responses in physics involves much more than checking final answers. Recent large language models (LLMs) make this problem testable. Vision-capable language models can read handwritten work, compare it with a rubric, and produce comments. Automated scoring predates generative AI, and reviews of short-answer and text-based assessment document a transition from hand-engineered features to learned language representations \cite{burrows2015,gao2024}. Earlier work showed that AI can grade responses to introductory physics problems at a useful level \cite{Kortemeyer23}. An independent study of university-level physics problems likewise found promising aggregate grading performance, but also question- and model-dependent errors \cite{Mok_2025}. A more demanding study on a high-stakes handwritten thermodynamics exam found that AI could support grading, while diagrams were harder than derivations and final grading still required human review \cite{Kortemeyer24}. A follow-up study argued that AI could handle some confident cases while humans review uncertain cases \cite{Kortemeyer25}. Similar conclusions appear in mathematics. Recent studies of handwritten calculus and university mathematics report strong agreement when OCR (optical character recognition), rubrics, and human verification are controlled \cite{liu2024,Kortemeyer25calculus,vanho26}. Studies of calculus submissions and national mathematics examinations also show strong automated-scoring performance alongside remaining item-level limitations~\cite{gandolfi2025,morris2025}. Work on handwritten graphs and mathematical OCR also shows that visual representation remains a separate difficulty beyond text recognition \cite{parsa25,nath25,seong26}. Broader benchmarks such as MathVista make the same point for mathematical reasoning in visual contexts \cite{lu2024}. Recent studies also show that automatic-scoring performance depends materially on the prompt design, rubric and item context, and choice of model~\cite{latif2024,lee2024,jiang2024,pecuchova2025}.

These studies show real progress and leave open an important question: Can AI grading help in handwritten physics examinations where the purpose includes a high-stakes outcome such as identifying a cohort or selecting a team? Physics Olympiad exam submissions are a useful test case because small score differences can affect candidates' rankings, selection outcomes, and medal awards. Olympiad questions are long and often require judgment beyond matching a textbook answer. A correct final answer may be reached for the wrong reason. An incorrect final answer may still show a sound method. Human examiners therefore judge both the answer and the route taken to reach it.

This study tests AI grading across a deliberately mixed set of physics assessments. OE1 is a national Olympiad theory examination used to identify a top cohort for the next stage from about $10\,000$ first-stage participants. OE2 is a final Olympiad selection camp comprising challenging theoretical and experimental examinations spanning several areas of high school physics and used to select the final team for the International Physics Olympiad. QM is an end-of-semester university examination in quantum mechanics and quantum computation. Together these assessments encompass the full spectrum of work physics examiners evaluate: theory, experiment, derivation, data analysis, diagrams, and conceptual reasoning.

The central question is how closely AI scores agree with official examiner scores, and whether the agreement is reliable enough for the relevant assessment outcome. Established automated-scoring frameworks treat validity as a property of the proposed use of the scores. The analysis must therefore include the decision consequences of the assessment system along with average agreement with a reference score \cite{williamson2012framework,kane2013validating,bennett1998validity}. A second question is how significantly the result depends on the details of how the grading instructions are written. The rubrics in this study were already detailed, including partial-credit rules, common errors, alternative valid approaches, and carried-forward errors. This rubric detail was central to the strong agreement reported here.

AI grading was done after the exams were completed and after the Olympiad selection results and university grades had already been released. The released official results remained unchanged by this retrospective AI analysis. After describing the assessments and grading procedure, the analysis evaluates agreement at the total-score, selection, question-part, and question-type levels. It then examines representative examples, experimental grading, and recurring disagreements before assessing rubric refinement and confidence-based human-review strategies. The two grading rounds are compared within those results, but the primary focus is agreement with official grading for handwritten physics exam submissions and the conditions under which AI may become useful in future grading.

\section{Data and methods}
\subsection{Assessments and reference scores}
All exam submissions in this study were handwritten. OE1 and OE2 come from different years and stages of the Indian Physics Olympiad program. For context, this program is a multistage selection process. In the first stage, typically $40\,000$--$50\,000$ students participate nationally, though participation was lower in the year of the OE1 examination used here ($10\,292$ students). About 350 students are then selected for the next written Olympiad stage, represented here by OE1. From this group, about 35--40 students are selected for the final Olympiad camp, represented here by OE2, and the final team of five students is selected from that camp.

The OE1 and OE2 datasets used in this study are from different years because of submission availability and scanning history. They therefore test AI grading at two different high-stakes stages of the same selection system. QM consists of 40 handwritten exam submissions from an end-of-semester university examination in quantum mechanics. In total, $10\,364$ scanned pages were processed for AI grading. Table~\ref{tab:datasets} summarizes the datasets.

\begin{table}[H]
\centering
\caption{Datasets used in the study. An exam component means one separately graded OE2 theory or experiment test section. An exam submission means one student's answer booklet for one component. OE1 and QM each had one component per student, while OE2 candidates completed five components.}
\label{tab:datasets}
\scriptsize
\begin{tabular}{@{}lrrrrl@{}}
\toprule
Dataset & \makecell{Candidates/\\students} & \makecell{Exam\\submissions} & \makecell{Exam\\components} & \makecell{Scanned\\pages} & Main role \\
\midrule
OE1 & 350 & 350 & 1 & 5600 & Top-cohort identification \\
OE2 & 26 & 130 & 5 & 4164 & Final team selection \\
QM  & 40 & 40 & 1 & 600 & University class grade \\
\midrule
Total & 416 & 520 & 7 & 10\,364 & -- \\
\bottomrule
\end{tabular}
\end{table}

The human examiner scores used here are the final official scores for each exam submission. Each question was graded by one examiner and checked by a second examiner before release. When the checking identified a concern, the mark was reconsidered and settled before scores were released. Students also had review mechanisms: OE1 allows regrading requests after scores are released, while in OE2 and QM students can inspect their graded submissions and discuss the points awarded with examiners. We therefore treat the final official scores, after these checks, as the reference scores for this study. Differences between official scores and AI scores are reported as disagreements with the official score. Because OE1 and OE2 are selection-oriented assessments, rank and top-group agreement are part of the grading question.

For OE2, the official final ranking followed the International Physics Olympiad theory--experiment weighting: theory contributes 30 marks and experiment contributes 20 marks, a 60:40 ratio. The two theory components were each originally out of 80 and were prorated to 120, so theory contributed 240 marks and the three experiment components contributed 160 marks, giving a combined score out of 400. In normalized analyses this is equivalent to combining the theory and experiment totals as \(0.6T+0.4E\).

\subsection{AI grading procedure and grading rounds}
The AI grading was performed both in browser-based runs and through Codex using the OpenAI API.  The AI grader was given the exam questions, solutions, official rubrics, and anonymized handwritten exam submissions. Visible human scores and comments on the submissions were erased before AI grading. The same official rubrics were used for human and AI grading. For human examiners, the rubrics served as scoring guidelines to be interpreted case by case, including responses that did not fit a listed solution path. The rubrics specified partial credit, common errors, alternative valid solution paths, and error-carried-forward rules. The AI prompt asked for these rules to be applied to the full written solution, including reasoning and intermediate steps, and to avoid repeated penalties for the same carried-forward error. The aim was to follow the International Olympiad exam grading standards.
Table~\ref{tab:modelruns} summarizes the model and effort settings used in the reported analyses.  The main comparison between the two full grading rounds uses GPT-5.5 Thinking at high effort. The 5.5 Pro runs were used to check whether the OE2 outcome was stable and to review selected exam submissions; the main OE2 selection outcome was unchanged in these checks.  The analysis is organized around  two full grading rounds and one focused refinement stage explained below:

\begin{itemize}
\item \textbf{Round I (RI):} the initial full AI grading. RI had no access to human scores, official totals, ranks, or selection status. The AI was provided the exam submissions, exam questions, solutions, rubrics, and the general grading instructions described above. The output included scores and comments for each question part or subpart.

\item \textbf{Focused refinement:} carried out after inspecting Round I score differences and AI comments. Three places with repeated or large disagreements, one each from QM, OE1, and OE2, were tested with more explicit instructions. These diagnostic runs tested whether stating the intended physics and scoring conditions more clearly could reduce AI-official score differences. The examples are discussed in Sec.~\ref{sec:refinement}.

\item \textbf{Round II (RII):} a new full grading run using instructions revised after the Round I disagreement analysis. During RII, the AI did not see RI scores, RI comments, human scores, official totals, ranks, selection status, or any AI--human comparison. It differed from RI in two main ways. First, it included the refined rubrics for the three parts discussed in the focused refinement stage. The remaining parts used the same rubrics as RI. Second, the AI was asked to grade long submissions page by page. A single submission in our exam set could exceed 30 scanned pages. RII therefore asked the AI to identify where the credited evidence appeared. The output also recorded confidence labels and review flags, discussed in Sec.~\ref{sec:confidence}.
\end{itemize}

\begin{table}[H]
\centering
\caption{AI models and effort settings used in the study.}
\label{tab:modelruns}
\begin{tabular}{@{}lll@{}}
\toprule
Model & Effort & Exams/runs included \\
\midrule
GPT-5.5 Thinking & High & Main RI/RII grading: OE1, OE2, QM \\
GPT-5.5 Pro & Standard & OE2 full comparison and outcome check \\
GPT-5.5 Pro & Extended & OE2 pilot involving selected submissions \\
\bottomrule
\end{tabular}
\end{table}

\subsection{Evaluation measures}
For analyses of individual official question parts, one comparison means one student response to one official part or subpart of a question. In these exams, questions were divided into parts and subparts, each with its own maximum mark; the grading scheme then specified how marks within that part were awarded or deducted. This gives 7058 comparisons between AI and official scores across OE1, OE2, and QM. Table~\ref{tab:metrics} defines the quantities used below.

\begin{table}[H]
\centering
\caption{Analysis parameters used in the study. Unless raw marks are explicitly stated, $D$ and MAD are reported as percentages of the relevant maximum possible score.}
\label{tab:metrics}
\begin{tabular}{@{}lll@{}}
\toprule
Quantity & What it measures & Purpose \\
\midrule
$r$ & \tablecell[0.39\textwidth]{Pearson correlation between human and AI scores} & \tablecell[0.32\textwidth]{Score tracking} \\
$\rho$ & \tablecell[0.39\textwidth]{Spearman rank correlation} & \tablecell[0.32\textwidth]{Rank-order agreement} \\
$D$ & \tablecell[0.39\textwidth]{Mean of AI minus human scores. Positive values mean AI awarded more} & \tablecell[0.32\textwidth]{Direction of over- or under-awarding} \\
MAD & \tablecell[0.39\textwidth]{Mean absolute difference between AI and human scores} & \tablecell[0.32\textwidth]{Size of the grading difference} \\
Top-$k$ overlap & \tablecell[0.39\textwidth]{Fraction of human top-$k$ submissions also in the AI top-$k$} & \tablecell[0.32\textwidth]{Top-group and team-selection checks} \\
$d=|\mathrm{AI}-\mathrm{Human}|$ & \tablecell[0.39\textwidth]{Raw point difference for one official question part} & \tablecell[0.32\textwidth]{Difference bands for individual question parts} \\~&&\\
Confidence flag & \tablecell[0.39\textwidth]{AI's self-reported confidence or request for human review} & \tablecell[0.32\textwidth]{Prioritizing human review in RII} \\
\bottomrule
\end{tabular}
\end{table}

\section{Agreement and selection results}
\subsection{Total-score agreement}
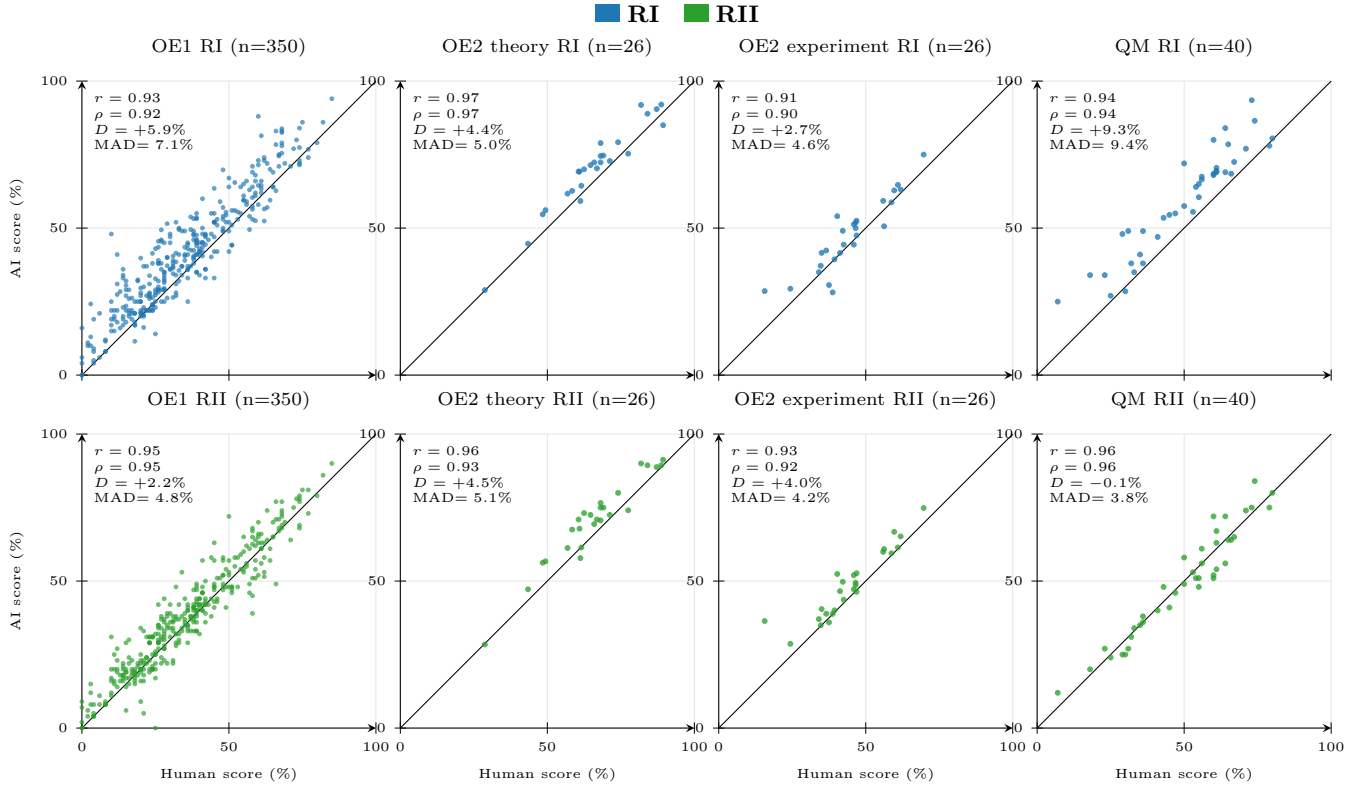
\begin{figure}[H]\centering
\barlegenditem{mainblue}{RI}\quad
\barlegenditem{maingreen}{RII}\par\vspace{0.2em}
\begin{tikzpicture}
\begin{groupplot}[
  group style={group size=4 by 2,horizontal sep=0.018\linewidth,vertical sep=0.78cm},
  totalscatteraxis
]
\nextgroupplot[title={OE1 RI (n=350)},ylabel={AI score (\%)},xticklabels={,,}]
\addplot[black,thin] coordinates {(0,0) (100,100)};
\addplot[only marks,mark=*,mark size=0.7pt,draw=mainblue,fill=mainblue,opacity=0.68]
  table[x=human_pct,y=ai_pct,col sep=comma]{data_tikz_total_oe1_RI.csv};
\scatterstats{0.93}{0.92}{+5.9}{7.1}

\nextgroupplot[title={OE2 theory RI (n=26)},xticklabels={,,}]
\addplot[black,thin] coordinates {(0,0) (100,100)};
\addplot[only marks,mark=*,mark size=0.9pt,draw=mainblue,fill=mainblue,opacity=0.78]
  table[x=human_pct,y=ai_pct,col sep=comma]{data_tikz_total_oe2_theory_RI.csv};
\scatterstats{0.97}{0.97}{+4.4}{5.0}

\nextgroupplot[title={OE2 experiment RI (n=26)},xticklabels={,,}]
\addplot[black,thin] coordinates {(0,0) (100,100)};
\addplot[only marks,mark=*,mark size=0.9pt,draw=mainblue,fill=mainblue,opacity=0.78]
  table[x=human_pct,y=ai_pct,col sep=comma]{data_tikz_total_oe2_experiment_RI.csv};
\scatterstats{0.91}{0.90}{+2.7}{4.6}

\nextgroupplot[title={QM RI (n=40)},xticklabels={,,}]
\addplot[black,thin] coordinates {(0,0) (100,100)};
\addplot[only marks,mark=*,mark size=0.9pt,draw=mainblue,fill=mainblue,opacity=0.78]
  table[x=human_pct,y=ai_pct,col sep=comma]{data_tikz_total_qm_RI.csv};
\scatterstats{0.94}{0.94}{+9.3}{9.4}

\nextgroupplot[title={OE1 RII (n=350)},xlabel={Human score (\%)},ylabel={AI score (\%)}]
\addplot[black,thin] coordinates {(0,0) (100,100)};
\addplot[only marks,mark=*,mark size=0.7pt,draw=maingreen,fill=maingreen,opacity=0.68]
  table[x=human_pct,y=ai_pct,col sep=comma]{data_tikz_total_oe1_RII.csv};
\scatterstats{0.95}{0.95}{+2.2}{4.8}

\nextgroupplot[title={OE2 theory RII (n=26)},xlabel={Human score (\%)}]
\addplot[black,thin] coordinates {(0,0) (100,100)};
\addplot[only marks,mark=*,mark size=0.9pt,draw=maingreen,fill=maingreen,opacity=0.78]
  table[x=human_pct,y=ai_pct,col sep=comma]{data_tikz_total_oe2_theory_RII.csv};
\scatterstats{0.96}{0.93}{+4.5}{5.1}

\nextgroupplot[title={OE2 experiment RII (n=26)},xlabel={Human score (\%)}]
\addplot[black,thin] coordinates {(0,0) (100,100)};
\addplot[only marks,mark=*,mark size=0.9pt,draw=maingreen,fill=maingreen,opacity=0.78]
  table[x=human_pct,y=ai_pct,col sep=comma]{data_tikz_total_oe2_experiment_RII.csv};
\scatterstats{0.93}{0.92}{+4.0}{4.2}

\nextgroupplot[title={QM RII (n=40)},xlabel={Human score (\%)}]
\addplot[black,thin] coordinates {(0,0) (100,100)};
\addplot[only marks,mark=*,mark size=0.9pt,draw=maingreen,fill=maingreen,opacity=0.78]
  table[x=human_pct,y=ai_pct,col sep=comma]{data_tikz_total_qm_RII.csv};
\scatterstats{0.96}{0.96}{-0.1}{3.8}
\end{groupplot}
\end{tikzpicture}
\caption{AI versus human total scores in RI and RII, expressed as percentages of the assessment maximum. Each point represents one student. The diagonal line is perfect agreement. Blue points show RI and green points show RII. $r$, $\rho$, $D$, and MAD are shown inside each panel.}
\label{fig:totalscatter}
\end{figure}

At the level of total scores, AI and official human scores track each other strongly across all three assessments. Fig.~\ref{fig:totalscatter} shows scatter plots of AI score against human score for OE1, QM, and the OE2 theory and experiment components. Each point represents one student. The points lie close to the equal-score diagonal in both rounds, although RII is generally less shifted toward positive AI--human differences than RI. 

The Pearson correlations are high in RI ($r=0.91$--$0.97$) and remain high in RII ($r=0.93$--$0.96$). RII was the full revised workflow described above: page-by-page checking, evidence notes, stricter checking of permitted scores, confidence/review fields, and clearer instructions for selected questions. The three focused refinements discussed later were worth only about 3.6\% of the combined official rubrics, so RI--RII differences should be interpreted as effects of the full revised workflow rather than those three questions alone. With detailed rubrics and instructions, current multimodal AI can reproduce the overall score distribution of handwritten physics assessments reasonably well.

Strong total-score agreement still leaves grading differences. In RI, $D$ was positive in all four total-score comparisons, meaning that AI generally awarded more points than human examiners. RII reduced this positive shift in OE1 and QM, as seen from the smaller $D$ values reported in Fig.~\ref{fig:totalscatter}. MAD gives the size of the remaining score difference. The clearest reductions in total-score MAD occurred for OE1 and QM: OE1 MAD fell from 7.1\% to 4.8\%, and QM MAD fell from 9.4\% to 3.8\%. OE2 already had strong total-score agreement in RI, and the RI--RII changes are smaller because OE2 combines several theoretical and experimental examinations. Similar over-awarding concerns have been reported in earlier AI-grading studies \cite{Kortemeyer24,Kortemeyer25}.

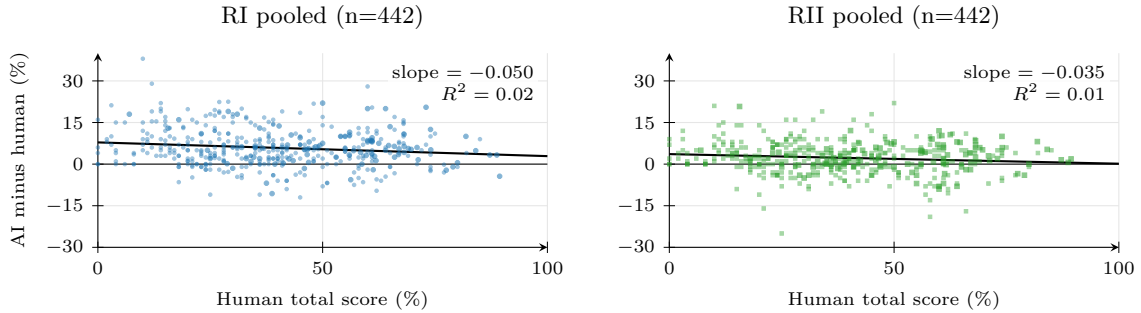
\begin{figure}[H]\centering
\begin{tikzpicture}
\begin{groupplot}[
  group style={group size=2 by 1,horizontal sep=0.09\linewidth},
  width=0.42\linewidth,height=4.15cm,
  xmin=0,xmax=100,ymin=-30,ymax=40,
  axis lines=left,
  axis line style={black},
  tick style={black},
  tick label style={font=\scriptsize,black},
  label style={font=\scriptsize,black},
  title style={font=\small,black},
  grid=major,
  grid style={gray!20},
  xtick={0,50,100},
  ytick={-30,-15,0,15,30},
  xlabel={Human total score (\%)},
  ylabel={AI minus human (\%)}
]
\nextgroupplot[title={RI pooled (n=442)}]
\addplot[black,thin,forget plot] coordinates {(0,0) (100,0)};
\addplot[only marks,mark=*,mark size=0.65pt,draw=mainblue,fill=mainblue,opacity=0.42]
  table[x=human_pct,y expr=\thisrow{ai_pct}-\thisrow{human_pct},col sep=comma]{data_tikz_total_oe1_RI.csv};
\addplot[only marks,mark=*,mark size=0.85pt,draw=mainblue,fill=mainblue,opacity=0.55]
  table[x=human_pct,y expr=\thisrow{ai_pct}-\thisrow{human_pct},col sep=comma]{data_tikz_total_oe2_theory_RI.csv};
\addplot[only marks,mark=*,mark size=0.85pt,draw=mainblue,fill=mainblue,opacity=0.55]
  table[x=human_pct,y expr=\thisrow{ai_pct}-\thisrow{human_pct},col sep=comma]{data_tikz_total_oe2_experiment_RI.csv};
\addplot[only marks,mark=*,mark size=0.85pt,draw=mainblue,fill=mainblue,opacity=0.55]
  table[x=human_pct,y expr=\thisrow{ai_pct}-\thisrow{human_pct},col sep=comma]{data_tikz_total_qm_RI.csv};
\addplot[black,thick,domain=0:100,samples=2] {7.83 - 0.0495*x};
\node[anchor=north east,align=right,font=\scriptsize,fill=white,fill opacity=0.82,text opacity=1,inner sep=1.5pt]
  at (rel axis cs:0.98,0.96) {slope \(=-0.050\)\\\(R^2=0.02\)};

\nextgroupplot[title={RII pooled (n=442)},ylabel={}]
\addplot[black,thin,forget plot] coordinates {(0,0) (100,0)};
\addplot[only marks,mark=square*,mark size=0.65pt,draw=maingreen,fill=maingreen,opacity=0.42]
  table[x=human_pct,y expr=\thisrow{ai_pct}-\thisrow{human_pct},col sep=comma]{data_tikz_total_oe1_RII.csv};
\addplot[only marks,mark=square*,mark size=0.85pt,draw=maingreen,fill=maingreen,opacity=0.55]
  table[x=human_pct,y expr=\thisrow{ai_pct}-\thisrow{human_pct},col sep=comma]{data_tikz_total_oe2_theory_RII.csv};
\addplot[only marks,mark=square*,mark size=0.85pt,draw=maingreen,fill=maingreen,opacity=0.55]
  table[x=human_pct,y expr=\thisrow{ai_pct}-\thisrow{human_pct},col sep=comma]{data_tikz_total_oe2_experiment_RII.csv};
\addplot[only marks,mark=square*,mark size=0.85pt,draw=maingreen,fill=maingreen,opacity=0.55]
  table[x=human_pct,y expr=\thisrow{ai_pct}-\thisrow{human_pct},col sep=comma]{data_tikz_total_qm_RII.csv};
\addplot[black,thick,domain=0:100,samples=2] {3.61 - 0.0346*x};
\node[anchor=north east,align=right,font=\scriptsize,fill=white,fill opacity=0.82,text opacity=1,inner sep=1.5pt]
  at (rel axis cs:0.98,0.96) {slope \(=-0.035\)\\\(R^2=0.01\)};
\end{groupplot}
\end{tikzpicture}
\caption{Residual total-score plots pooled across OE1, OE2 theory, OE2 experiment, and QM. The vertical axis is AI minus human total score, expressed as a percentage of the assessment maximum. Blue circles show RI and green squares show RII. The fitted line is a least-squares visual guide, and the horizontal zero line is perfect agreement. The figure mainly shows the direction and spread of AI--human differences: RI is more often positive, while RII is less shifted upward. The fitted slopes should be read cautiously because the human score is part of the plotted difference.}
\label{fig:residualscore}
\end{figure}

To check the direction and spread of the differences, Fig.~\ref{fig:residualscore} plots AI minus human score against the human total score, pooled across the four total-score comparisons. Positive values mean AI awarded more than the human examiner. RI points are more often above zero, while RII reduces this upward shift. The differences appear across the score range rather than only among low-scoring submissions or near a cutoff. The fitted lines are included as a visual guide; because the vertical axis already contains the human score, the slope should not be over-interpreted as showing that AI is better or worse for high-scoring students.

\subsection{Top-cohort and selection agreement}
For assessments that classify or select candidates, average-score agreement is only one part of the evaluation. The practical question is whether AI also matches the relevant outcome. OE1 identifies a cohort for the next stage, OE2 selects a final team of five students, and QM assigns a course grade. Classifications based on test scores have their own accuracy and consistency properties \cite{livingston1995}, so these outcome checks are part of the grading question.

For OE1, AI recovered most of the larger top group. Each round recovered 31 of the human top 40 and 40 of the human top 50. At smaller cutoffs RII improved the overlap. For example, the top-10 overlap increased from 3/10 in RI to 7/10 in RII. This level of agreement is useful for identifying the top group, while human grading remains necessary for exact ranks near a cutoff.

For OE2, the key question is whether AI matches the selected group for India's IPhO (International Physics Olympiad) team. In both RI and RII, the AI top five contained the same five students as the human top five, although the AI ranked them in a different order. This comparison is based on five exam components per candidate, so the final ranking combines more evidence than a single examination. At $k=10$, both rounds recovered 9 of the human top 10, and at $k=20$ both recovered 19 of the human top 20. At the OE2 top-five cutoff, overlap was complete in both rounds.

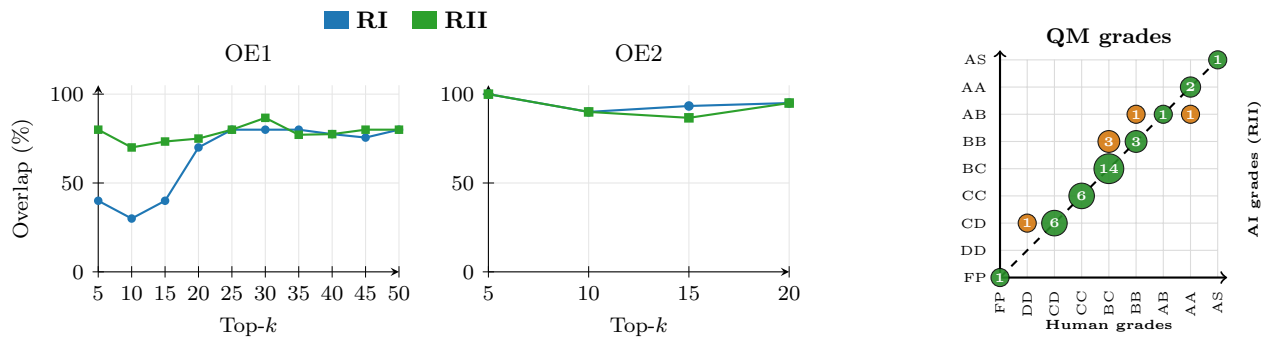
\begin{figure}[H]\centering
\begin{minipage}[t]{0.66\linewidth}
\centering
\barlegenditem{mainblue}{RI}\quad
\barlegenditem{maingreen}{RII}\par\vspace{0.2em}
\begin{tikzpicture}
\begin{groupplot}[
  group style={group size=2 by 1,horizontal sep=0.10\linewidth},
  width=0.47\linewidth,height=4.05cm,
  axis lines=left,
  axis line style={black},
  tick style={black},
  tick label style={font=\footnotesize,black},
  label style={font=\footnotesize,black},
  title style={font=\small,black},
  ymin=0,ymax=105,
  grid=major,
  grid style={gray!20},
  xlabel={Top-$k$},
  ylabel={Overlap (\%)}
]
\nextgroupplot[title={OE1},xtick={5,10,15,20,25,30,35,40,45,50}]
\addplot[mainblue,mark=*,mark size=1.2pt,thick] table[x=k,y=RI,col sep=comma]{data_tikz_topk_oe1.csv};
\addplot[maingreen,mark=square*,mark size=1.2pt,thick] table[x=k,y=RII,col sep=comma]{data_tikz_topk_oe1.csv};

\nextgroupplot[title={OE2},xtick={5,10,15,20},ylabel={}]
\addplot[mainblue,mark=*,mark size=1.4pt,thick] table[x=k,y=RI,col sep=comma]{data_tikz_topk_oe2.csv};
\addplot[maingreen,mark=square*,mark size=1.4pt,thick] table[x=k,y=RII,col sep=comma]{data_tikz_topk_oe2.csv};
\end{groupplot}
\end{tikzpicture}
\end{minipage}\hfill
\begin{minipage}[t]{0.30\linewidth}
\centering
\vspace{0.35em}
\begin{tikzpicture}[x=0.36cm,y=0.36cm]
  \draw[step=1,gray!28,thin] (0,0) grid (8,8);
  \draw[black,thick,dashed] (0,0) -- (8,8);
  \draw[black,thick,->] (0,0) -- (8.35,0);
  \draw[black,thick,->] (0,0) -- (0,8.35);
  \foreach \x/\lab in {0/FP,1/DD,2/CD,3/CC,4/BC,5/BB,6/AB,7/AA,8/AS}{
    \node[font=\tiny,rotate=90,anchor=east] at (\x,-0.16) {\lab};
    \node[font=\tiny,anchor=east] at (-0.16,\x) {\lab};
  }
  \node[font=\tiny\bfseries] at (4,-1.78) {Human grades};
  \node[font=\tiny\bfseries,rotate=90] at (9.35,4) {AI grades (RII)};
  \node[font=\footnotesize\bfseries] at (4,8.82) {QM grades};
  \foreach \x/\y/\n/\r in {
    0/0/1/3.4,
    2/2/6/4.8,
    3/3/6/4.8,
    4/4/14/5.6,
    5/5/3/4.1,
    6/6/1/3.4,
    7/7/2/3.8,
    8/8/1/3.4
  }{
    \filldraw[fill=maingreen!88!black,draw=black,line width=0.4pt,opacity=0.90] (\x,\y) circle[radius=\r pt];
    \node[font=\tiny\bfseries,text=white] at (\x,\y) {\n};
  }
  \foreach \x/\y/\n/\r in {
    1/2/1/3.4,
    4/5/3/4.1,
    5/6/1/3.4,
    7/6/1/3.4
  }{
    \filldraw[fill=mainorange!95!black,draw=black,line width=0.4pt,opacity=0.92] (\x,\y) circle[radius=\r pt];
    \node[font=\tiny\bfseries,text=white] at (\x,\y) {\n};
  }
\end{tikzpicture}
\end{minipage}
\caption{Outcome agreement. The OE1 and OE2 panels show the percentage of human top-$k$ submissions or students also present in the AI top-$k$ group; blue circles show RI and green squares show RII. The QM panel shows RII course-grade agreement across the nine released grade categories. Green bubbles show exact grade matches and orange bubbles show different grade pairs. The number inside each bubble is the number of students; AS is the highest grade.}
\label{fig:topcohort}
\end{figure}

QM has a different outcome: a course grade. The released grades span nine categories, from FP to AS. We therefore compared the AI-based course grades with the released official grades. RI exactly matched 26 of 40 official grades, while RII exactly matched 34 of 40, and all 40 RII grades were within one grade step on this nine-category scale (Fig.~\ref{fig:topcohort}). AS is the highest grade.

\FloatBarrier

The OE2 outcome was also checked for dependence on AI mode. The Pro Standard run gave a similar result: it recovered the same human top-five group as the main Thinking High run, although the internal order changed. At broader cutoffs, Pro Standard recovered 9 of the human top 10 and 19 of the human top 20. A Pro Extended pilot was run only on six selected OE2 submissions with large earlier AI--human disagreements. It modestly reduced some disagreements, especially in theory, while the same pattern of over-awarding remained. Although the setting and model configuration differ from ours, earlier work has also reported nontrivial run-to-run variation when AI graded the same responses \cite{jauhiainen}. Overall, the OE2 team configuration was stable across these AI-mode checks. This mode comparison was limited to OE2 and was not repeated for OE1 or QM.

\subsection{Question-part agreement and partial credit}
Many questions in the exams were divided into parts and subparts. Scores on these official question parts therefore give a stricter test than total scores. In Table~\ref{tab:bands}, each official question part is compared separately, using $d=|\mathrm{AI}-\mathrm{Human}|$ in raw points. The usual scoring increment was 0.5, so the table separates exact matches, differences up to 0.5 points, differences between 0.5 and 1 point, and differences larger than 1 point. Because the maximum mark differs across parts, the \(d\) bands should be read together with the normalized MAD values and the zero/partial/full-credit analysis below.

Across all 7058 official question parts, RI matched the human score exactly in about 63\% of parts. RII increased exact agreement to about 70\%, and reduced parts differing by more than one point from about 13\% to 7\%.

\begin{table}[H]
\centering
\caption{Agreement between AI and human scores for individual question parts. Here $d=|\mathrm{AI}-\mathrm{Human}|$ in raw points. Values are rounded percentages. Bands sum to 100\% before rounding.}
\label{tab:bands}
\scriptsize
\setlength{\tabcolsep}{5.5pt}
\begin{tabular}{l r @{\quad} rr @{\quad} rr @{\quad} rr @{\quad} rr}
\toprule
& & \multicolumn{2}{c}{$d=0$} & \multicolumn{2}{c}{$0<d\leq0.5$} & \multicolumn{2}{c}{$0.5<d\leq1$} & \multicolumn{2}{c}{$d>1$} \\
\cmidrule(lr){3-4}\cmidrule(lr){5-6}\cmidrule(lr){7-8}\cmidrule(lr){9-10}
Set & Question parts & RI & RII & RI & RII & RI & RII & RI & RII \\
\midrule
OE1 & 4200 & 61 & 71 & 15 & 12 & 9 & 9 & 15 & 8 \\
OE2 theory & 1482 & 70 & 75 & 13 & 13 & 9 & 6 & 8 & 6 \\
OE2 experiment & 416 & 45 & 56 & 21 & 17 & 9 & 13 & 25 & 14 \\
QM & 960 & 63 & 69 & 21 & 18 & 8 & 9 & 8 & 4 \\
All official question parts & 7058 & 63 & 70 & 16 & 14 & 9 & 9 & 13 & 7 \\
\bottomrule
\end{tabular}
\par\smallskip
\parbox{0.96\linewidth}{\scriptsize The OE2 experiment row reports the official experiment question parts. Because some official experiment parts combine several judgments into one mark, RII was also checked at a more detailed experimental-component level for analysis only. In that check (Sec.~\ref{sec:fineexp}), exact agreement was 77\%, with 16\% in \(0<d\leq0.5\), 4\% in \(0.5<d\leq1\), and 3\% in \(d>1\).}
\end{table}

\begin{figure}[H]
\centering
\resizebox{0.92\linewidth}{!}{%
\begin{tikzpicture}[x=0.078cm,y=0.42cm]
\node[font=\scriptsize] at (36,7.80) {{\color{mainblue}\rule{0.62em}{0.62em}} AI = zero};
\node[font=\scriptsize] at (62,7.80) {{\color{mainorange}\rule{0.62em}{0.62em}} AI = partial};
\node[font=\scriptsize] at (87,7.80) {{\color{maingreen}\rule{0.62em}{0.62em}} AI = full};

\node[anchor=east,font=\scriptsize\bfseries] at (-18,6.15) {Human = zero ($n=2870$)};
\node[anchor=east,font=\scriptsize\bfseries] at (-18,3.70) {Human = partial ($n=1834$)};
\node[anchor=east,font=\scriptsize\bfseries] at (-18,1.25) {Human = full ($n=2354$)};

\node[anchor=east,font=\scriptsize] at (-4,6.65) {RI};
\node[anchor=east,font=\scriptsize] at (-4,5.65) {RII};
\node[anchor=east,font=\scriptsize] at (-4,4.20) {RI};
\node[anchor=east,font=\scriptsize] at (-4,3.20) {RII};
\node[anchor=east,font=\scriptsize] at (-4,1.75) {RI};
\node[anchor=east,font=\scriptsize] at (-4,0.75) {RII};

\draw[fill=mainblue,draw=black,line width=0.2pt] (0,6.28) rectangle (69.27,7.02);
\draw[fill=mainorange,draw=black,line width=0.2pt] (69.27,6.28) rectangle (97.77,7.02);
\draw[fill=maingreen,draw=black,line width=0.2pt] (97.77,6.28) rectangle (100,7.02);
\node[font=\scriptsize,text=white] at (34.64,6.65) {69\%};
\node[font=\scriptsize,text=white] at (83.52,6.65) {29\%};

\draw[fill=mainblue,draw=black,line width=0.2pt] (0,5.28) rectangle (80.35,6.02);
\draw[fill=mainorange,draw=black,line width=0.2pt] (80.35,5.28) rectangle (98.61,6.02);
\draw[fill=maingreen,draw=black,line width=0.2pt] (98.61,5.28) rectangle (100,6.02);
\node[font=\scriptsize,text=white] at (40.18,5.65) {80\%};
\node[font=\scriptsize,text=white] at (89.48,5.65) {18\%};

\draw[fill=mainblue,draw=black,line width=0.2pt] (0,3.83) rectangle (1.25,4.57);
\draw[fill=mainorange,draw=black,line width=0.2pt] (1.25,3.83) rectangle (83.80,4.57);
\draw[fill=maingreen,draw=black,line width=0.2pt] (83.80,3.83) rectangle (100,4.57);
\node[font=\scriptsize,text=white] at (42.53,4.20) {83\%};
\node[font=\scriptsize,text=white] at (91.90,4.20) {16\%};

\draw[fill=mainblue,draw=black,line width=0.2pt] (0,2.83) rectangle (1.96,3.57);
\draw[fill=mainorange,draw=black,line width=0.2pt] (1.96,2.83) rectangle (88.44,3.57);
\draw[fill=maingreen,draw=black,line width=0.2pt] (88.44,2.83) rectangle (100,3.57);
\node[font=\scriptsize,text=white] at (45.20,3.20) {86\%};
\node[font=\scriptsize,text=white] at (94.22,3.20) {12\%};

\draw[fill=mainblue,draw=black,line width=0.2pt] (0,1.38) rectangle (0.59,2.12);
\draw[fill=mainorange,draw=black,line width=0.2pt] (0.59,1.38) rectangle (16.18,2.12);
\draw[fill=maingreen,draw=black,line width=0.2pt] (16.18,1.38) rectangle (100,2.12);
\node[font=\scriptsize,text=white] at (8.39,1.75) {16\%};
\node[font=\scriptsize,text=white] at (58.09,1.75) {84\%};

\draw[fill=mainblue,draw=black,line width=0.2pt] (0,0.38) rectangle (1.87,1.12);
\draw[fill=mainorange,draw=black,line width=0.2pt] (1.87,0.38) rectangle (12.62,1.12);
\draw[fill=maingreen,draw=black,line width=0.2pt] (12.62,0.38) rectangle (100,1.12);
\node[font=\scriptsize,text=white] at (7.25,0.75) {11\%};
\node[font=\scriptsize,text=white] at (56.31,0.75) {87\%};
\end{tikzpicture}}
\caption{AI behavior by official human score group. Each bar shows the distribution of AI scores within question parts where the official human score was zero, partial credit, or full credit. The partial-credit parts contain 40\% of the available points.}
\label{fig:zeropartialfullbehavior}
\end{figure}
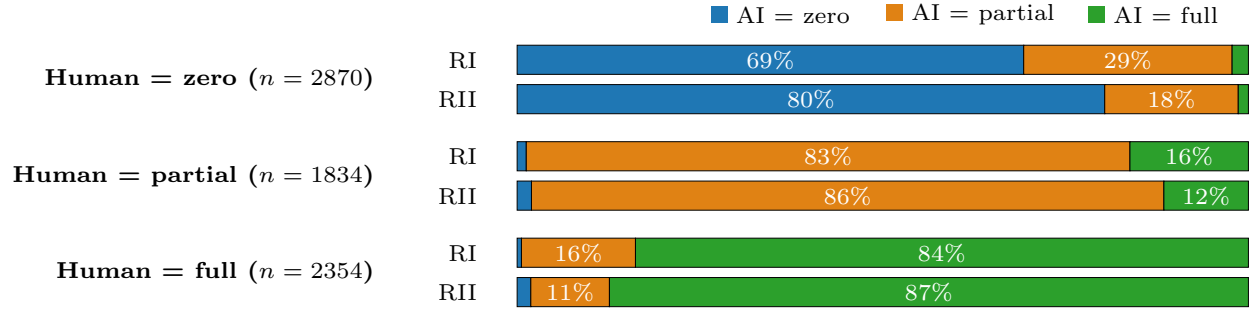

Separating question parts by the official human score gives a more useful view of overall exact agreement. Fig.~\ref{fig:zeropartialfullbehavior} shows two levels of agreement. At the broad level, AI often placed responses in the correct zero-, partial-, or full-credit band. In RII, it kept 80\% of human-zero parts at zero, 86\% of human-partial parts in the partial-credit range, and 87\% of human-full parts at full credit. This band-level agreement matters because partial-credit parts account for 40\% of the available points.

The harder task is the exact amount of partial credit. For human-partial parts, exact agreement rose from 24.6\% in RI to 32.8\% in RII. Among cases where both the human examiner and AI gave partial credit, MAD fell from 0.92 to 0.72 raw marks. When AI moved a human-zero part into the partial-credit range, the average award was about one raw mark, and this happened less often in RII than RI. Thus, RII improved both broad category recognition and the calibration of partial credit, while exact partial-credit scoring remained the most difficult case.

To understand what lies behind these aggregate question-part results, we next examine representative successes, experimental grading, differences across question types, and recurring disagreement patterns.

\section{Strengths and disagreements in AI grading}
\subsection{Examples of physics-specific grading}
In many cases the AI comments were aligned with the physics conditions in the rubric. The model often found the relevant solution in lengthy, multipage exam submissions, even when the final summary answer and the detailed work appeared on different pages. It interpreted handwritten derivations, credited equivalent mathematical forms, and sometimes identified specific physics errors. These examples show that the total-score agreement was supported by physics-specific grading comments, with the AI often identifying the evidence relevant to the score.

\begin{figure}[H]
\centering
\begin{minipage}[b]{0.38\linewidth}
\centering
\includegraphics[width=\linewidth]{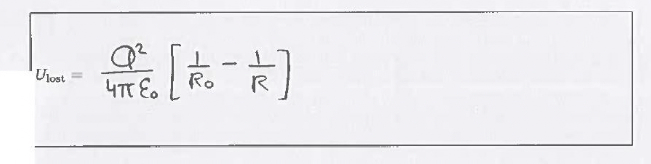}
\par\smallskip\textbf{(a)}
\end{minipage}\hfill
\begin{minipage}[b]{0.58\linewidth}
\centering
\includegraphics[width=\linewidth]{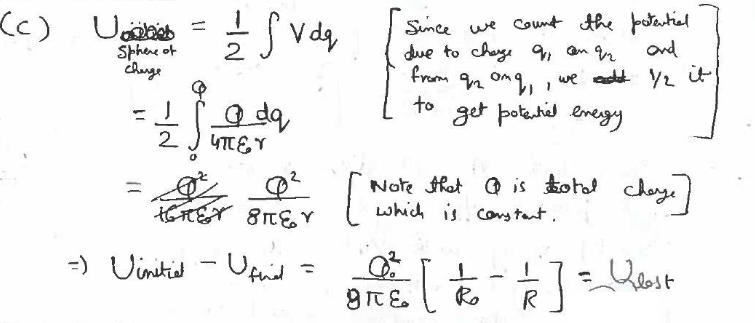}
\par\smallskip\textbf{(b)}
\end{minipage}
\caption{Example of AI using the full working to interpret a summary-box transcription error. (a) The transferred expression in the summary answer box is off by a factor of two. (b) The detailed working of the same submission contains the correct self-energy factor and final expression. The AI comment stated: ``\textit{The summary box misses a factor, but the working sheet correctly derives $Q_0^2/(8\pi\epsilon_0)(1/R_0 - 1/R)$.}'' This matched the official full-credit mark.}
\label{fig:positivefactor}
\end{figure}

One example is a transcription error between the working pages and the summary answer box. In Fig.~\ref{fig:positivefactor}, the copied final expression in the summary box is off by a factor of two, but the detailed working contains the correct self-energy factor and the correct final result. The AI credited the detailed working despite the transcription error in the summary box. This matters in Olympiad-style grading, where students often copy final results into a summary answer space and a copying error is judged in the context of the full submission.

\begin{figure}[H]
\centering
\begin{minipage}{0.30\linewidth}
\centering
\includegraphics[width=\linewidth]{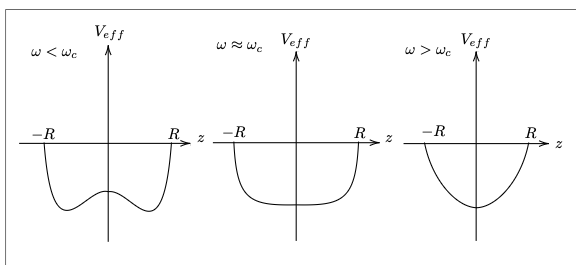}
\par\smallskip\textbf{(a)}
\end{minipage}\hfill
\begin{minipage}{0.32\linewidth}
\centering
\includegraphics[width=\linewidth]{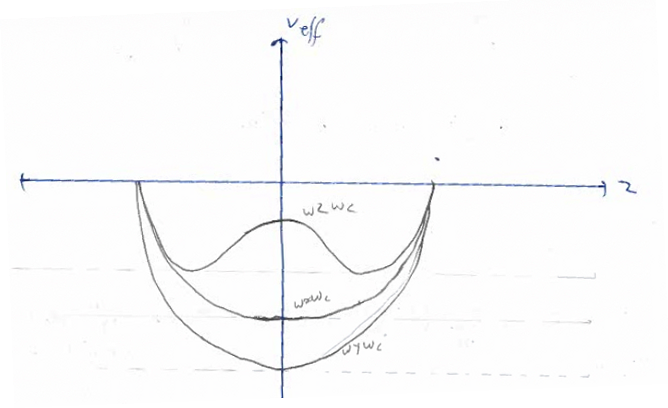}
\par\smallskip\textbf{(b)}
\end{minipage}\hfill
\begin{minipage}{0.32\linewidth}
\centering
\includegraphics[width=\linewidth]{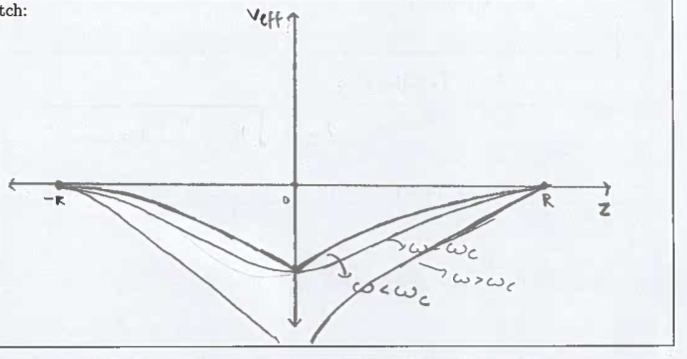}
\par\smallskip\textbf{(c)}
\end{minipage}
\caption{Examples of visual reasoning in AI grading. (a) Correct reference shapes from the solution. (b) Full-credit example where the relevant sketches were outside the designated answer box. AI found them on later pages and commented: ``\textit{Detailed sketches show the required below-critical, critical, and above-critical effective-potential shapes with the expected qualitative behavior.}'' (c) Limited-credit example where AI awarded the endpoint behavior specified in the rubric but withheld credit for the missing double-well and critical-transition shapes. The AI comment stated: ``\textit{Provides some qualitative potential curves and endpoint behavior, but the subcritical double-well and critical transition are not represented correctly.}''}
\label{fig:positivediagram}
\end{figure}

A related example involved visual grading. The AI was often able to read qualitative sketches and identify physically relevant features. In one question, examinees were required to draw the effective-potential diagram, \(V_{\rm eff}\) versus \(z\). In one submission, the required sketches appeared on later pages outside the designated answer box. The AI found them and awarded full credit. In another case, it correctly gave only limited credit because the sketch showed curves meeting the zero line at the endpoints, but lacked the double-well and critical shapes required by the rubric (Fig.~\ref{fig:positivediagram}, with exact AI comments in the caption). This kind of interpretation matters in Olympiad grading because small score differences can affect ranking.

\FloatBarrier

\subsection{Experimental grading}
\label{sec:fineexp}
The OE2 experimental submissions are a useful case because they require reading tables, calculations, graphs, fit lines, uncertainty estimates, and written conclusions. Some official experiment parts were broad, with several of these judgments contributing to the same mark. RII also recorded them more separately, for example at the level of table quality, graphing, fit or slope extraction, uncertainty, and conclusion. These experimental components were compared with the corresponding human markings recorded in the official grading scheme. This more detailed experimental check is summarized below Table~\ref{tab:bands}.

The AI often recognized transformed tables, plotted points, fit and limiting lines, slopes, and reported values. The weaker cases were those where the score depended on whether the data range, fit, uncertainty, and final conclusion supported each other.

\begin{figure}[H]
\centering
\begin{minipage}[t]{0.32\linewidth}
\centering
\includegraphics[width=\linewidth,height=4.25cm,keepaspectratio]{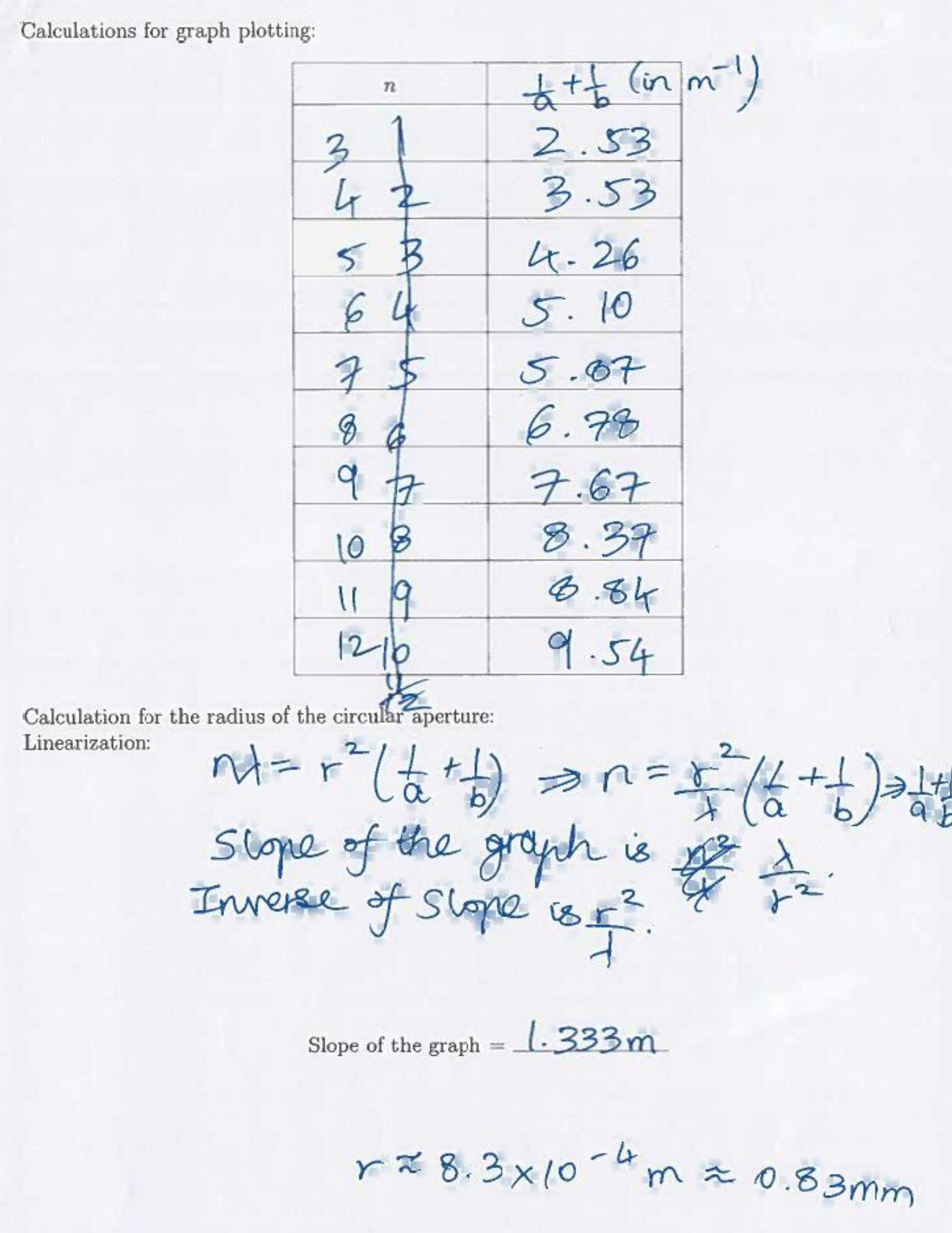}
\par\smallskip\textbf{(a)}
\end{minipage}\hfill
\begin{minipage}[t]{0.32\linewidth}
\centering
\includegraphics[width=\linewidth,height=4.25cm,keepaspectratio]{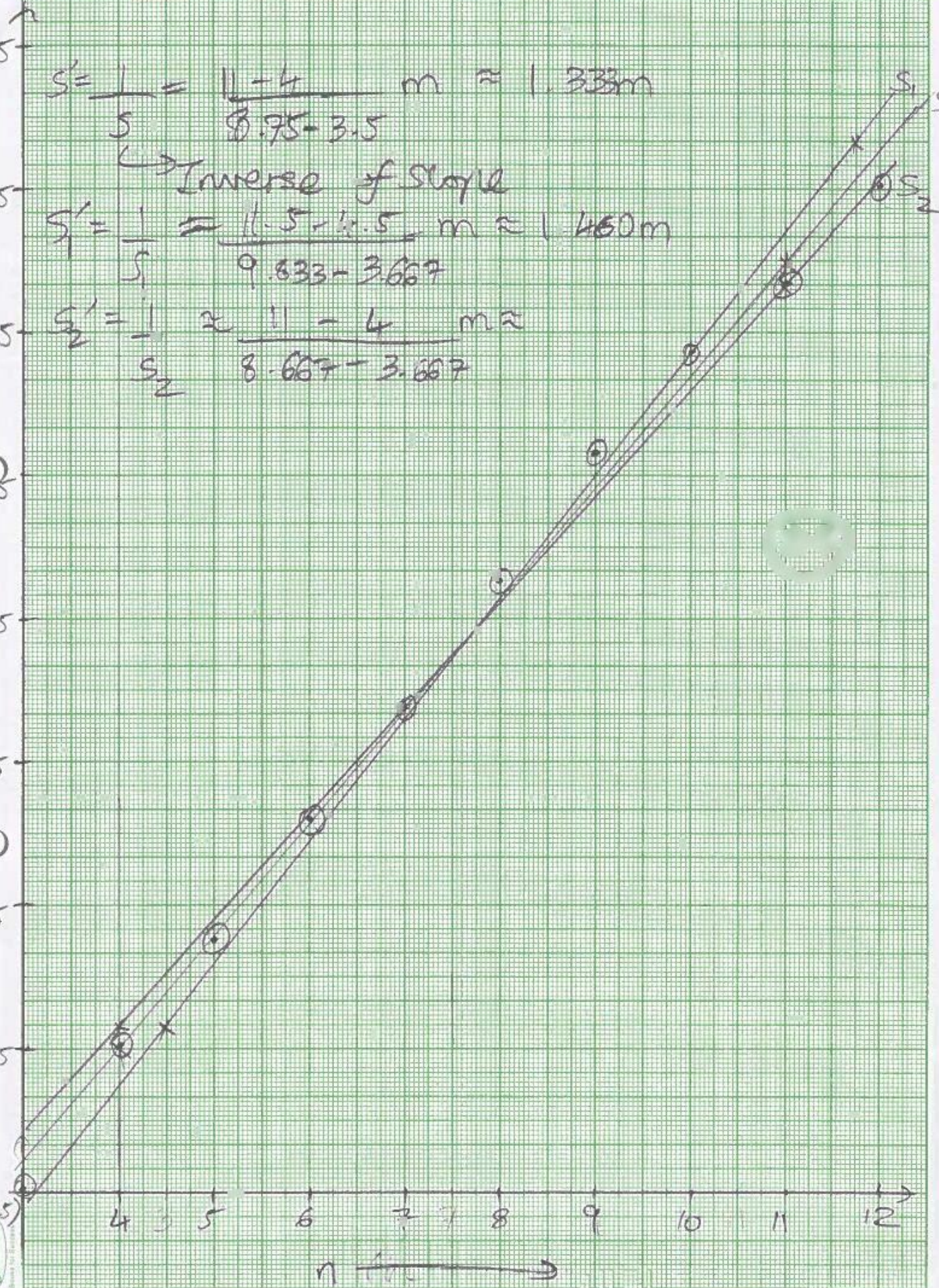}
\par\smallskip\textbf{(b)}
\end{minipage}\hfill
\begin{minipage}[t]{0.32\linewidth}
\centering
\includegraphics[width=\linewidth,height=4.25cm,keepaspectratio]{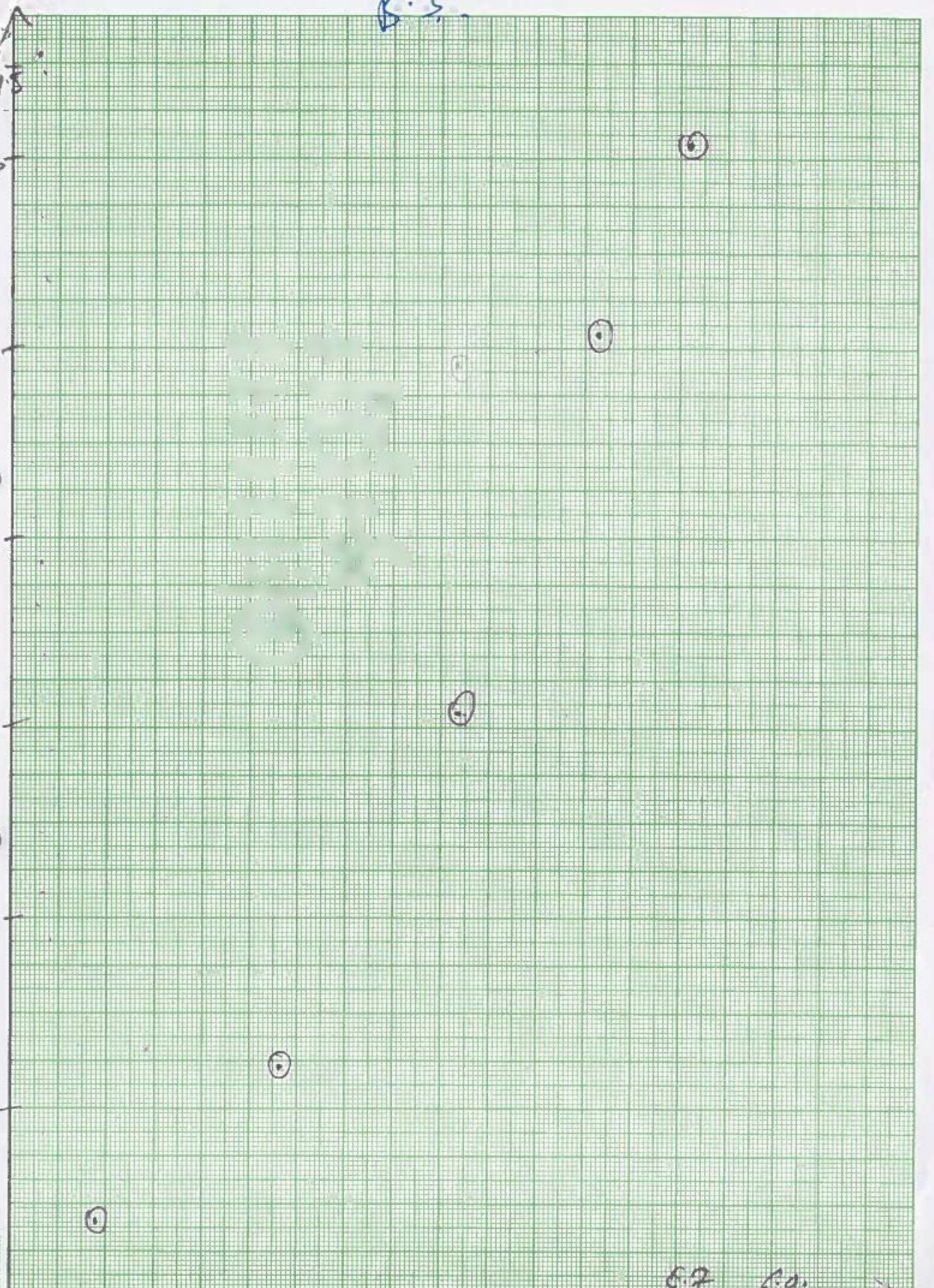}
\par\smallskip\textbf{(c)}
\end{minipage}
\caption{Examples from masked OE2 experimental submissions. Panels (a) and (b) belong to the same data-table and graph-analysis task, and the AI and official scores agreed on both marking components. {AI comments for these two components were:} ``\textit{The transformed $1/a+1/b$ values are tabulated for all $n$ values and the linearization is written}'' and ``\textit{Graph is present with plotted points, best-fit and limiting lines, and slope calculations for the equivalent $1/a+1/b$ plot.}'' Panel (c) shows a partial-credit graph. Both graders awarded partial credit. {AI comment for this component was:} ``\textit{Five points are plotted on an $R$-versus-$\sqrt{h}$ graph, but no usable best-fit line, slope or uncertainty construction is shown.}''}
\label{fig:experimentalexamples}
\end{figure}

The examples in Fig.~\ref{fig:experimentalexamples} show that the scoring went beyond graph presence. AI identified fit and limiting lines when they were part of the grading evidence, and it agreed with partial credit when plotted points alone fell short of the full graph-analysis requirement.

These examples indicate that multimodal AI can often read and score experimental evidence. Human review remains important when the mark depends on whether the table, graph, fit, uncertainty, and conclusion are physically consistent with each other.

\FloatBarrier
\subsection{Agreement across question types}
To compare OE1, OE2, and QM using the same labels, each official question part was assigned to one primary question type. These labels are broad, but they capture the main type of judgment the grader has to make (Table~\ref{tab:categories}). For OE2 experimental sections, the label refers to the main task in the official question part.

\begin{table}[H]
\centering
\caption{Question-type labels used across the three assessments.}
\label{tab:categories}
\begin{tabular}{@{}ll@{}}
\toprule
\tablecell[0.25\textwidth]{Question type} & \tablecell[0.65\textwidth]{Description} \\
\midrule
\tablecell[0.25\textwidth]{Conceptual written reasoning} & \tablecell[0.65\textwidth]{Written physical explanation, evaluation of an answer choice, interpretation, or validity check.} \\
\tablecell[0.25\textwidth]{Figure-diagram} & \tablecell[0.65\textwidth]{Circuits, free-body diagrams, qualitative sketches, ray diagrams, or graph shapes where the visual structure is the answer.} \\
\tablecell[0.25\textwidth]{Numerical} & \tablecell[0.65\textwidth]{Calculation or substitution where the quantity being assessed is primarily a value, probability, time, energy, or angle.} \\
\tablecell[0.25\textwidth]{Derivation} & \tablecell[0.65\textwidth]{Symbolic proof, formula derivation, operator derivation, or multi-step algebraic reasoning.} \\
\tablecell[0.25\textwidth]{Figure-data} & \tablecell[0.65\textwidth]{Experimental measurements, data tables, plotted data, graph fitting, slope extraction, uncertainty, or data-based inference.} \\
\bottomrule
\end{tabular}
\end{table}

This question-type analysis shows that agreement extends beyond numerical work. In derivations, numerical answers, conceptual reasoning, diagrams, and data/graph tasks, AI scores generally rise when human scores rise. The differences lie in how large the score differences are on individual parts. Conceptual written reasoning shows the greatest over-scoring in RI, while figure-data and figure-diagram tasks are sensitive to whether the visual or experimental evidence satisfies the specific grading condition.

\begin{figure}[H]\centering
\barlegenditem{mainblue}{RI}\quad
\barlegenditem{maingreen}{RII}\par\vspace{0.2em}
\begin{tikzpicture}
\begin{axis}[
  paperbaraxis,
  width=0.62\linewidth,height=4.0cm,
  symbolic x coords={Conceptual,Diagram,Numerical,Derivation,Experiment},
  xtick=data,
  xticklabels={Conceptual,Diagram,Numerical,Derivation,Fig.-data},
  x tick label style={rotate=18,anchor=east,font=\scriptsize},
  ymin=0,ymax=19,
  ylabel={MAD\%}]
\addplot[fill=mainblue,draw=mainblue] table[x=Category,y=RIMAD,col sep=comma]{data_tikz_category_mad.csv};
\addplot[fill=maingreen,draw=maingreen] table[x=Category,y=RIIMAD,col sep=comma]{data_tikz_category_mad.csv};
\end{axis}
\end{tikzpicture}
\caption{MAD by question type in RI and RII for the 7058 official question parts. MAD\% is the mean absolute difference as a percentage of the maximum possible score for each part.}
\label{fig:categorymetrics}
\end{figure}
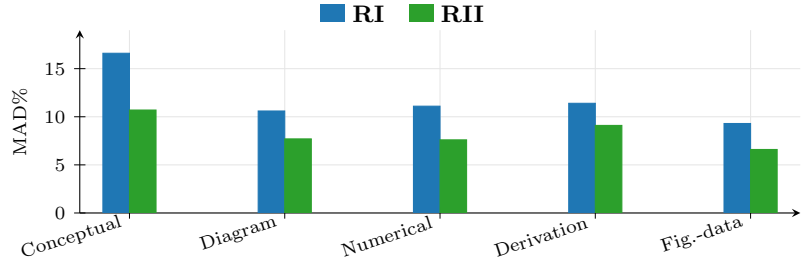

Fig.~\ref{fig:categorymetrics} and Table~\ref{tab:catmetrics} summarize the question-type results. MAD is lower in RII for every question type. The reviewed cases suggest why: page-by-page checking and clearer credit conditions help most when the score depends on evidence in a specific part of the student's response.

\begin{table}[H]
\centering
\caption{Question-type metrics for official question parts. $N$ is the number of question-part comparisons. $D$ and MAD are percentages of the maximum possible score for each part.}
\label{tab:catmetrics}
\begin{tabular}{lrrrrrrr}
\toprule
Question type & $N$ & $D_{\mathrm{I}}$ & $D_{\mathrm{II}}$ & MAD$_{\mathrm{I}}$ & MAD$_{\mathrm{II}}$ & $r_{\mathrm{I}}$ & $r_{\mathrm{II}}$ \\
\midrule
Conceptual written reasoning & 706 & +13.3 & +4.3 & 16.6 & 10.7 & 0.83 & 0.89 \\
Figure-diagram & 1082 & +3.8 & +2.5 & 10.6 & 7.7 & 0.84 & 0.89 \\
Numerical & 1978 & +4.7 & +0.7 & 11.1 & 7.6 & 0.85 & 0.88 \\
Derivation & 2876 & +4.8 & +2.3 & 11.4 & 9.1 & 0.88 & 0.91 \\
Figure-data & 416 & +3.3 & +4.4 & 9.3 & 6.6 & 0.92 & 0.96 \\
\bottomrule
\end{tabular}
\end{table}

The question-type results also show that RII gains extended beyond the explicitly refined questions. Figure-data and figure-diagram question parts contain several of the largest improvements, but numerical, conceptual, and derivation parts improved as well. The central concern is whether the table, graph, diagram, or derivation satisfies the specific physical criterion required by the rubric.

\FloatBarrier
\subsection{Recurring sources of disagreement}
The larger disagreements usually had identifiable causes, summarized in Table~\ref{tab:failuremodes}. These patterns motivated the focused refinement tests and the later revised full grading.

\begin{table}[H]
\centering
\caption{Recurring sources of disagreement between AI and official scores. These summarize patterns seen in reviewed cases; individual score differences are interpreted against the official scoring.}
\label{tab:failuremodes}
\begin{tabular}{@{}ll@{}}
\toprule
\tablecell[0.28\textwidth]{Source} & \tablecell{Description} \\
\midrule
\tablecell[0.28\textwidth]{Awarding credit too easily} & \tablecell{AI sometimes awarded partial credit to work to which human examiners assigned zero points, especially when the answer contained plausible symbols, text, tables, or diagram features.} \\
\tablecell[0.28\textwidth]{Over-valuing the final answer} & \tablecell{A correct-looking final answer sometimes received too much credit even when the reasoning had serious errors.} \\
\tablecell[0.28\textwidth]{Treatment of carried-forward errors} & \tablecell{AI could recognize an earlier error but sometimes applied carried-forward-error rules differently from the human examiner.} \\
\tablecell[0.28\textwidth]{Missing a required diagram feature} & \tablecell{AI often read the diagram but could miss a specific feature required by the rubric, such as relative placement, curvature, scale, or a required label.} \\
\tablecell[0.28\textwidth]{Experimental evidence chain} & \tablecell{AI often identified tables, graphs, and calculations, but disagreements arose when the score depended on whether the data, fit, uncertainty, and conclusion supported each other.} \\
\tablecell[0.28\textwidth]{Permitted scoring increments} & \tablecell{In a few cases, AI awarded increments smaller than the intended scoring resolution. This should be stated clearly in the prompt.} \\
\bottomrule
\end{tabular}
\end{table}

These patterns lead to the two checks below: clearer scoring conditions and confidence flags for human review.

\FloatBarrier
\section{Improving and reviewing AI grades}
\subsection{Focused rubric refinement}
\label{sec:refinement}
Reviewing RI disagreements identified cases where the rubric stated a broad goal while leaving the specific physics needed for credit implicit. A focused test was therefore run on three problem cases: QM question 3, part b, OE1 Q2, and OE2 T2 Q1(f) (Table~\ref{tab:refinementtargets}). Each was regraded with clearer question-specific instructions to test whether the AI-official score difference could be reduced. In these cases, making the required physics and scoring conditions explicit reduced several RI disagreements.

\begin{table}[H]
\centering
\caption{Focused refinement targets selected after reviewing RI disagreements.}
\label{tab:refinementtargets}
\begin{tabular}{@{}llll@{}}
\toprule
Target & Points & Responses & Main issue clarified \\
\midrule
QM Q3b & 3 & 40 & Grover diagram and operator conditions \\
OE1 Q2 & 8 & 72 & A thermodynamic reasoning based question \\
OE2 T2 Q1(f) & 4 & 26 & $T$--$S$ graph shape, labels, and curvature \\
\bottomrule
\end{tabular}
\end{table}

\begin{figure}[H]
\centering
\begin{minipage}{0.26\linewidth}
\centering
\includegraphics[width=\linewidth]{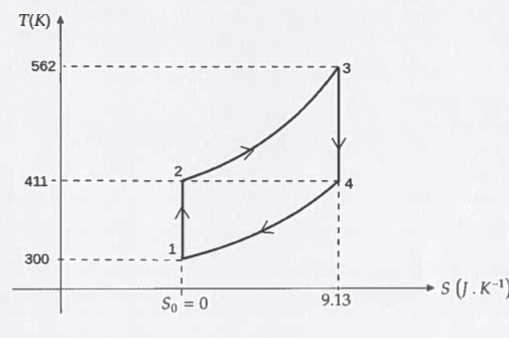}
\par\smallskip\textbf{(a)}
\end{minipage}\hspace{0.06\linewidth}
\begin{minipage}{0.26\linewidth}
\centering
\includegraphics[width=\linewidth]{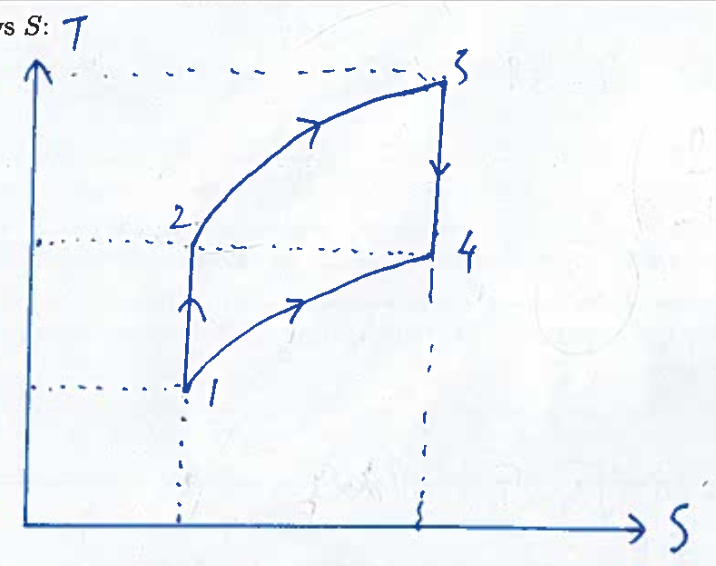}
\par\smallskip\textbf{(b)}
\end{minipage}
\caption{Refinement of the rubric for the OE2 question. (a) The intended $T$--$S$ cycle from the solution. (b) Student sketch with labels and arrows but with the wrong curvature. In RI, AI marked this as a correct shape. The refined rubric emphasized the physical conditions of the cycle, with less reliance on labels and arrows alone.}
\label{fig:tsgraph}
\end{figure}

In OE2, one part of a theory question asked students to draw a thermodynamic cycle on a $T$--$S$ plot (Fig.~\ref{fig:tsgraph}). The original rubric allotted 0.5 points for the correct shape, but left the intended curvature implicit. The AI often awarded the 0.5 points for the shape even if the overall shape or curvature was physically wrong, though it had access to the correct intended shape in model solutions. The refined instruction stated the conditions directly: ``The branches $1\to2$ and $3\to4$ had to be vertical isentropic branches. The branch $2\to3$ had to be an isobaric heating curve, convex upward on a $T$--$S$ plot with slope increasing left-to-right. The branch $4\to1$ had to be an isobaric cooling curve traversed right-to-left. The temperatures had to satisfy $T_3$ highest, $T_1$ lowest, and $T_2=T_4$.''

With the refined rubric, AI stopped awarding shape credit in the incorrect-curvature cases we had identified. For the submission in Fig.~\ref{fig:tsgraph}, the AI comment in this round said that labels and values were present, but ``the upper branch had the wrong curvature'' and one process arrow was inconsistent. 

\begin{figure}[H]
\centering
\begin{minipage}[t]{0.52\linewidth}
\vspace{0pt}
\centering
\begin{minipage}[t]{0.38\linewidth}
\vspace{0pt}
\centering
\resizebox{\linewidth}{!}{%
\begin{tikzpicture}[>=latex,scale=0.78]
  \coordinate (O) at (0,0);
  \draw[gray!45,thin] (O) circle (2.0);
  \draw[->,thick] (-0.12,0) -- (2.50,0) node[right,font=\tiny] {$|x_0^\perp\rangle$};
  \draw[->,thick] (0,-0.12) -- (0,2.35) node[above,font=\tiny] {$|x_0\rangle$};
  \draw[->,black!80,very thick] (O) -- (24:1.95) node[above right,font=\tiny,xshift=1pt] {$|S\rangle$};
  \draw[->,black!55,thick] (O) -- (114:1.95) node[above left,font=\tiny] {$|S'\rangle$};
  \draw[->,mainblue,very thick] (O) -- (12:1.85) node[right,font=\tiny] {$|\psi\rangle$};
  \draw[->,gray!75,thick,densely dashed] (O) -- (-12:1.85) node[right,font=\tiny] {$O|\psi\rangle$};
  \draw[->,maingreen!80!black,very thick] (O) -- (60:1.90) node[above,font=\tiny] {$DO|\psi\rangle$};
  \draw[->,thin] (0.92,0) arc[start angle=0,end angle=24,radius=0.92];
  \node[font=\tiny] at (13:1.10) {$\theta$};
  \draw[->,thin] (12:1.18) arc[start angle=12,end angle=60,radius=1.18];
  \node[font=\tiny] at (38:1.38) {$2\theta$};
\end{tikzpicture}}
\end{minipage}\hfill
\begin{minipage}[t]{0.58\linewidth}
\vspace{0pt}
{\footnotesize
Choose $|x_0^\perp\rangle$ such that\\[0.2em]
\(\displaystyle |S\rangle=\cos\theta |x_0^\perp\rangle+\sin\theta |x_0\rangle.\)}
\end{minipage}
\par\vspace{0.25em}
{\footnotesize
\setlength{\abovedisplayskip}{2pt}
\setlength{\belowdisplayskip}{2pt}
Let $|\psi\rangle=\cos\alpha |x_0^\perp\rangle+\sin\alpha |x_0\rangle$, with $\alpha$ measured from $|x_0^\perp\rangle$. Then
\[
\begin{aligned}
O|\psi\rangle&=\cos\alpha |x_0^\perp\rangle-\sin\alpha |x_0\rangle,\\
DO|\psi\rangle&=\cos(\alpha+2\theta)|x_0^\perp\rangle\\
&\quad+\sin(\alpha+2\theta)|x_0\rangle .
\end{aligned}
\]
\par}
\end{minipage}\hspace{0.02\linewidth}
\begin{minipage}[t]{0.44\linewidth}
\vspace{0pt}
\centering
\includegraphics[width=\linewidth,trim={0 55pt 0 0},clip]{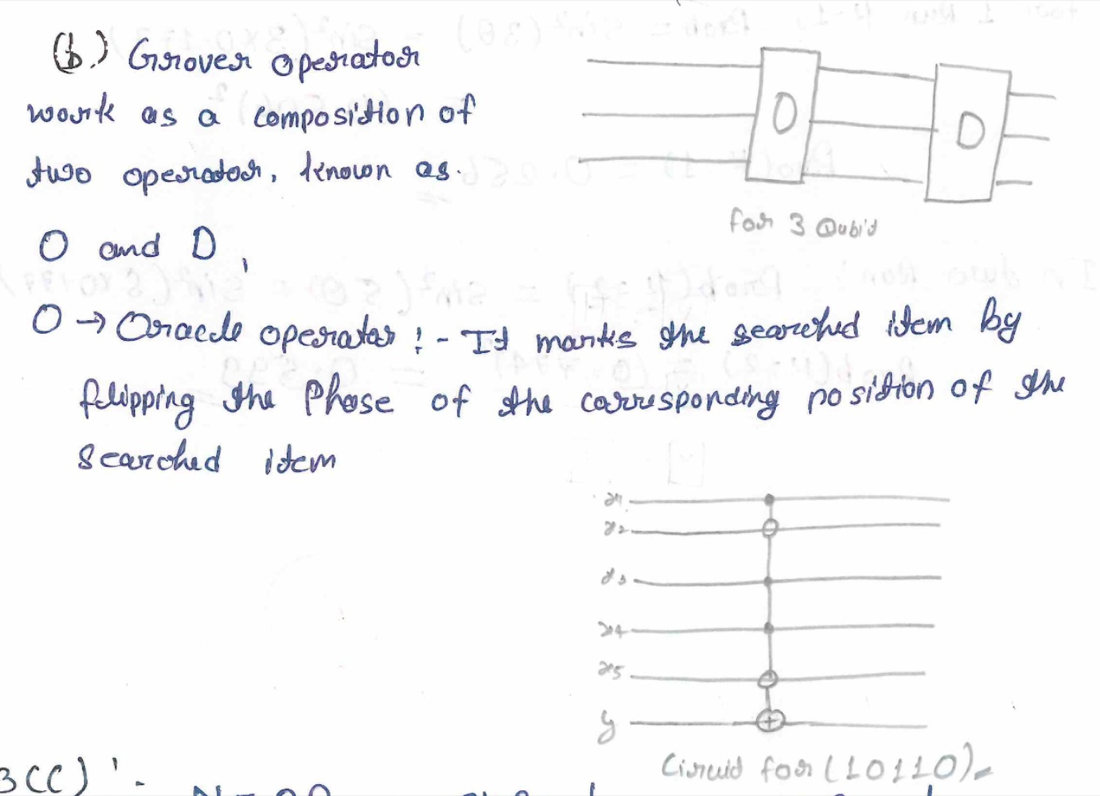}
\end{minipage}
\par\smallskip
\begin{minipage}[t]{0.52\linewidth}
\centering
\textbf{(a)}
\end{minipage}\hspace{0.02\linewidth}
\begin{minipage}[t]{0.44\linewidth}
\centering
\textbf{(b)}
\end{minipage}
\caption{Grover-rotation example from the QM exam. (a) Reference rotation diagram redrawn from the official solution. (b) Student response. In RII, the AI identified that the circuit-style sketch was not the required rotation diagram.}
\label{fig:groverexample}
\end{figure}
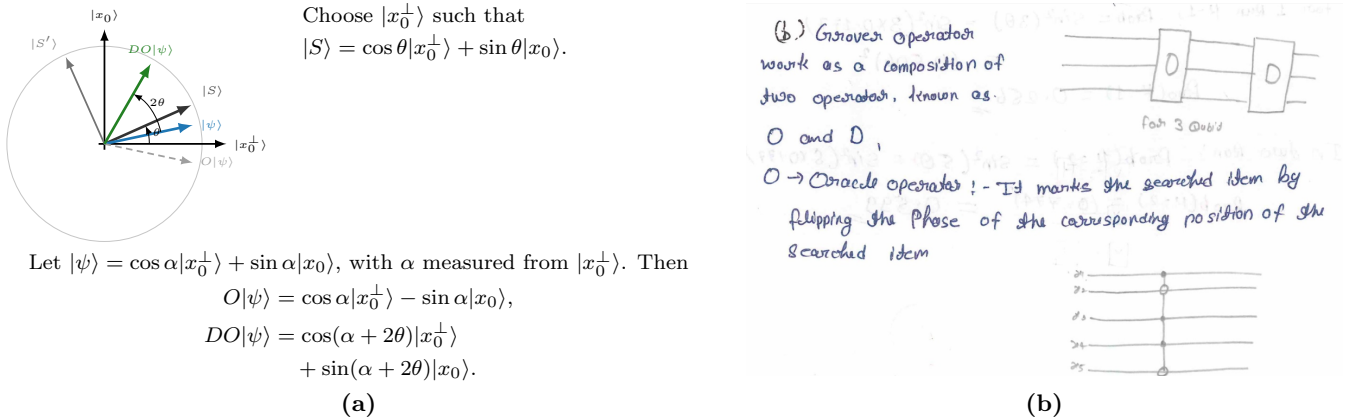

Another example is QM question 3, part b, which involved the Grover-rotation diagram. The original rubric provided only broad criteria for scoring the diagram and mathematical description. In several submissions, the AI awarded credit for a diagram that looked plausible but lacked the required elements. Fig.~\ref{fig:groverexample}(b) shows one such case. The response contained a circuit-style sketch and a qualitative oracle/diffusion description, but lacked the required two-dimensional rotation diagram and the operator action expected in the solution.

The rubric was then rewritten as explicit items (Table~\ref{tab:grover_scheme}). In this example, RI awarded 2 out of 3 points. RII matched the official 0.5 out of 3 points by crediting the qualitative oracle/diffusion idea and rejecting the sketch as the required Grover-rotation diagram.

\begin{table}[H]
\centering
\small
\caption{RI and RII rubric/comment comparison for the Grover example in Fig.~\ref{fig:groverexample}.}
\label{tab:grover_scheme}
\begin{tabular}{@{}lll@{}}
\toprule
\tablecell[0.11\textwidth]{} & \tablecell[0.39\textwidth]{\textbf{RI}} & \tablecell[0.39\textwidth]{\textbf{RII}} \\
\midrule
\tablecell[0.11\textwidth]{Rubric} &
\tablecell[0.39\textwidth]{Correct diagram with all symbols explained: 1 point. Detailed mathematical description of the Grover operator: 2 points. Determine an appropriate point allocation when the response is incomplete.} &
\tablecell[0.39\textwidth]{0.5 points for axes labeled as $|x_0\rangle$ and $|x_0^\perp\rangle$. 0.5 for showing rotation by $2\theta$ under $G$. 0.5 for defining $|x_0^\perp\rangle$. 0.5 for the action of $O$. 1 point for the action of $G=DO$.} \\
\addlinespace
\tablecell[0.11\textwidth]{Comment\\by AI} &
\tablecell[0.39\textwidth]{Awarded 2/3. \textit{Correctly describes Grover as oracle plus diffusion and notes the oracle phase flip, with a supporting circuit-style diagram. Deducted 1 because the diffusion/reflection and amplitude-rotation geometry are not fully explained. Method partly follows the official scheme.}} &
\tablecell[0.39\textwidth]{Awarded 0.5/3. \textit{Gives a qualitative oracle/diffusion description and a circuit-like oracle sketch, but not the required two-dimensional rotation diagram or mathematical action of $O$ and $G$ on $\cos(\alpha)|x_0^\perp\rangle+\sin(\alpha)|x_0\rangle$.}}   \\
\bottomrule
\end{tabular}
\end{table}

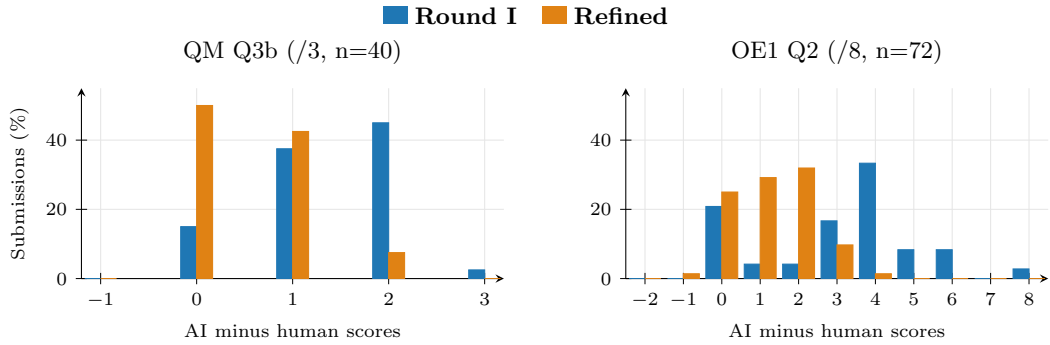
\begin{figure}[H]\centering
\barlegenditem{mainblue}{Round I}\quad
\barlegenditem{mainorange}{Refined}\par\vspace{0.2em}
\begin{tikzpicture}
\begin{groupplot}[
  group style={group size=2 by 1,horizontal sep=0.09\linewidth},
  width=0.40\linewidth,height=4.1cm,
  ybar=0pt,
  /pgf/bar width=6pt,
  axis lines=left,
  axis line style={black},
  tick style={black},
  tick label style={font=\scriptsize,black},
  label style={font=\scriptsize,black},
  title style={font=\small,black},
  ymin=0,ymax=55,
  ylabel={Submissions (\%)},
  xlabel={AI minus human scores},
  grid=major,
  grid style={gray!20},
  enlarge x limits=0.05,
]
\nextgroupplot[title={QM Q3b (/3, n=40)},xtick={-1,0,1,2,3}]
\addplot[fill=mainblue,draw=mainblue] table[x=x,y=Original,col sep=comma]{data_tikz_refinement_qm_q3b.csv};
\addplot[fill=mainorange,draw=mainorange] table[x=x,y=Refined,col sep=comma]{data_tikz_refinement_qm_q3b.csv};

\nextgroupplot[title={OE1 Q2 (/8, n=72)},xtick={-2,-1,0,1,2,3,4,5,6,7,8},ylabel={}]
\addplot[fill=mainblue,draw=mainblue] table[x=x,y=Original,col sep=comma]{data_tikz_refinement_oe1_q2.csv};
\addplot[fill=mainorange,draw=mainorange] table[x=x,y=Refined,col sep=comma]{data_tikz_refinement_oe1_q2.csv};
\end{groupplot}
\end{tikzpicture}
\caption{AI-minus-human score distributions for two focused refinement examples, shown as percentages of submissions. The horizontal axis gives the difference between AI and human scores. The Round I bars show the original scores, and the Refined bars show scores from the question-specific refinement run.}
\label{fig:refinementhist}
\end{figure}

For this question, the refinement reduced MAD from 1.1 to 0.4 marks, increased Pearson's $r$ from 0.70 to 0.82, raised exact agreement from 15\% to 45\%, and raised agreement within 0.5 mark, including exact matches, from 35\% to 80\%. Fig.~\ref{fig:refinementhist} shows the AI-minus-human score distributions for this case and the OE1 Q2 refinement example.

 OE1 Q2 was an 8-point conceptual thermodynamics question requiring reasoning for each option. Only the OE1 Q2 submissions with large Round I AI-official score differences ($n=72$) were regraded to test whether clearer wording reduced these disagreements. The refined rubric required the relevant physical criterion for each option and limited credit for unsupported conclusions. Within this subset, MAD fell from 38\% to 17\% of the maximum possible score for the part, and Pearson's $r$ increased from 0.58 to 0.83.

Taken together, the focused refinements support the practical requirement that reliable AI grading depends on a detailed, physics-specific marking rubric. Before AI grading is attempted, the rubric should state as explicitly as possible the conditions for awarding marks, common deductions, acceptable alternative solutions, and carried-forward-error rules.

\subsection{Confidence and human-review flags}
\label{sec:confidence}
\begin{figure}[H]\centering
\barlegenditem{maingreen}{High confidence}\quad
\barlegenditem{mainorange}{Medium/low confidence}\par\vspace{0.2em}
\begin{tikzpicture}
\begin{axis}[paperbaraxis,width=0.39\linewidth,height=4.0cm,
  symbolic x coords={OE1,OE2 Theory,OE2 Expt,QM},
  xtick={OE1,OE2 Theory,OE2 Expt,QM},
  x tick label style={align=center,font=\scriptsize},
  ymin=0,ymax=32,
  ylabel={MAD\%}]
\addplot[fill=maingreen,draw=maingreen]
  table[x=Set,y=HighMAD,col sep=comma]{data_confidence_mae.csv};
\addplot[fill=mainorange,draw=mainorange]
  table[x=Set,y=NonHighMAD,col sep=comma]{data_confidence_mae.csv};
\end{axis}
\end{tikzpicture}
\caption{Confidence flags in RII grading. Medium- and low-confidence question parts have larger differences from official scores, so confidence helps identify parts for human review.}
\label{fig:confidence}
\end{figure}
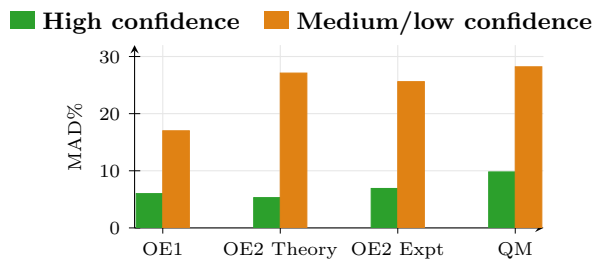

Confidence and human-review flags for each question part were available in RII. Fig.~\ref{fig:confidence} compares high-confidence question parts with medium- or low-confidence question parts. In every comparison where the flag was available, high-confidence parts had much lower MAD. For example, OE1 RII had MAD\% values of 6.0\% for high-confidence parts and 17.0\% for medium- or low-confidence parts. QM RII had 9.8\% versus 28.2\%. In OE2, theory parts showed 5.3\% versus 27.1\%, while experiment parts showed 6.9\% versus 25.6\%.

Confidence flags still miss some important disagreements. For the OE2 confidence analysis, the 1898 official question parts consisted of 1482 theory parts and 416 experiment parts. Accepting every high-confidence part with no review flag would have accepted 1724 parts; 87 of these still differed from the official human score by more than one point. These results support review triage as the appropriate use: confidence flags help decide what humans should inspect first, while final acceptance still needs examiner judgment. This conclusion aligns with prior work on model confidence and AI-assisted scoring \cite{tian2023,li2025}. It also resembles systems that send uncertain cases to a human expert; the present study evaluates the AI's own review flags, with no separately trained routing model \cite{mozannar2020}.
\FloatBarrier
\section{Discussion and conclusions}
The results show strong performance, but also clear limitations. AI grading showed strong agreement with the official total scores. For OE1, it identified most of the larger top group in both rounds. For OE2, both RI and RII recovered the same top-five group as the human examiners, although the AI ranked them differently. For QM, RII reproduced 34 of 40 released course grades exactly, and all 40 grades were within one grade step in a nine-step grading system. These findings support the use of current multimodal AI as a useful aid for physics grading in high-stakes examinations.

This study used moderated official scores as the human reference. These scores had already been checked by more than one examiner and were available for student review. A separate human--human regrading experiment was outside the scope of this analysis. Reported human-rater data for handwritten physics constructed responses show why this matters: for 20 physics constructed responses scored by four instructors, Tang, Ambrose, and Cheng reported human-only intraclass correlations (ICCs) of 0.88 with a holistic rubric and 0.94 with a checklist rubric, and only 0.28 for mid-level responses under the holistic rubric, rising to 0.89 with the checklist \cite{tang2026rubric}. Partial-credit scoring is therefore difficult for human examiners too, and explicit credit conditions are the shared remedy. Prior automated-scoring work treats human scoring evidence, and in some cases human--human agreement, as the benchmark for interpreting model performance \cite{williamson2012framework,morris2025}. A useful next step would be independent second-human grading on a representative subset of these same submissions, so that AI--official and human--official differences can be compared under the same conditions.

The question-part results explain where caution is still warranted. AI is often effective at reading handwritten work, recognizing correct physics, and identifying relevant evidence within large amounts of text and diagrams. It can credit correct working despite a transcription error in the summary box, accept equivalent forms, and identify important diagram features. It is also more willing than human examiners to award partial credit for incomplete or visually plausible work, particularly in conceptual reasoning, diagrams, and experimental graph or data tasks, where human examiners rely more heavily on context and judgment.

The refinement experiments show that useful prompt detail specifies the physics conditions required for awarding credit. This improved grading in the Grover diagram question in QM, where the diagram requirements had to be made explicit, and in OE1 Q2, where the expected reasoning for each option had to be clearly stated. The full RII results extend this pattern most clearly for OE1 and QM. OE2 total-score agreement was already high in RI, so its RI--RII changes were smaller and mixed, although question-part and experimental analyses still improved. Because the RII instructions were revised after inspecting RI disagreements in these submissions, the RII gains should be confirmed on an untouched examination or a holdout set. The OE2 results also demonstrate that total-score agreement alone can be misleading, since similar totals may conceal differences in the marks awarded to individual parts. Future prompts should specify the conditions for awarding marks explicitly.

Confidence and review fields provided a useful way to prioritize manual review. High-confidence responses showed smaller average score differences across all examinations. Some high-confidence responses still differed from the official human markings, so these measures are best used to prioritize review effort, with examiner checking before final acceptance.

A practical lesson from RII is that the unit of grading matters. Asking the AI to grade page by page and identify the evidence supporting each score appeared especially useful for long handwritten submissions, where relevant work may appear outside the expected answer space. This approach reduces the burden of processing an entire submission at once, improves auditability, and was associated with better agreement with human grading, although it also increased grading time. The human grading workload was also substantial. OE1 grading of about 350 submissions typically involves roughly 20 graders over about 2.5 days, with individual graders often working more than 12 hours per day. OE2 grading is a still larger distributed effort: about 25 people work intensively over four days, with the team operating nearly round the clock during final scoring and scrutiny. Because the study combined browser- and API-based workflows, comparable token usage and processing times were incompletely recorded. The study also lacked human grading-time logs suitable for direct comparison, so these staffing figures should be read as workload context rather than a measured time comparison. Future work should measure these costs directly.

The practical implication is that expert examiners remain central. Reliable agreement with official grading requires a detailed rubric that makes the award and deduction of marks as explicit as possible. AI shifts part of examiner effort toward defining the rubric carefully, stating the conditions for credit, and reviewing flagged or borderline cases. For practical use, the more intensive Pro modes fit best as targeted audit tools. The main Thinking High run already reproduced the OE2 top-five outcome, and the Pro Standard and limited Pro Extended checks left the grading picture largely unchanged. Pro modes may still be useful for targeted audits of difficult cases, and in this study their outcome-level performance was comparable to Thinking High. Overall, AI is best used as a second reader, an audit tool, and an aid to selection decisions. It can flag submissions with large differences from official scores, provide an additional set of comments, and help examiners check grading consistency.
\section*{Data and materials statement}
Anonymized scores by question part, task categories, prompts, and analysis scripts may be shared subject to approval from the relevant examination authorities and institutions. Raw handwritten submissions remain confidential because of student privacy and examination confidentiality.

\section*{Acknowledgments}
The authors thank the human examiners and the team that scanned the exam submissions. We thank Mamatha Maddur for helping to digitize and organize the submissions. P.P. acknowledges support from the Fulbright Program administered by the Institute of International Education (IIE). This work was partially performed under the auspices of the U.S. Department of Energy by Lawrence Livermore National Laboratory under Contract DE-AC52-07NA27344 (D.R.). The Physics Olympiad program in India is supported by the Government of India, Department of Atomic Energy, under project identification number RTI4001.

\end{document}